# Dimension-Dependent Critical Scaling Analysis and Emergent Competing Interaction Scales in a 2D Van der Waals magnet $Cr_2Ge_2Te_6$

P C Mahato[1], Suprotim Saha[1], Bikash Das[3], Subhadeep Datta[3], Rajib Mondal[5], Sourav Mal[4], Ashish Garg[2], Prasenjit Sen[4,*], S. S. Banerjee[1,+]

[1]*Department of Physics, Indian Institute of Technology Kanpur, Kanpur, Uttar Pradesh 208016, India*

[2] *Department of Sustainable Energy Engineering, Indian Institute of Technology Kanpur, Kanpur, Uttar Pradesh 208016, India*

[3]*School of Physical Sciences, Indian Association for the Cultivation of Science, Jadavpur, Kolkata -700032, India*

[4]*Harish-Chandra Research Institute, HBNI, Chhatnag Road, Jhunsi, Allahabad, Uttar Pradesh 211019, India*

[5]*UGC-DAE Consortium for Scientific Research, Kolkata Centre, Bidhannagar, Kolkata, 700106 India*

Corresponding authors email ID:

*: prasen@hri.res.in

+: satyajit@iitk.ac.in

**Abstract:**

We investigate thickness-dependent transformation from a paramagnetic to ferromagnetic phase in $Cr_2Ge_2Te_6$ (CGT) in bulk and few-layer flake forms. 2D Ising-like critical transition in bulk CGT occurs at $T_c$ = 67 K with out-of-plane magnetic anisotropy. Few-layer CGT on $hBN/SiO_2/Si$ substrate displays the same $T_c$ but also exhibits a new critical transition at $T_c'$ =14.2 K. Here, critical scaling analysis reveals the critical exponents differ significantly from those in bulk and do not align with the known universality classes. Our Density Functional Theory (DFT) and classical calculations indicate competition between magnetocrystalline and dipolar anisotropy emerges with reduced dimensions. The observed behavior is due to minor structural distortions in low dimensional CGT, which modify the balance between spin-orbit coupling, exchange interactions and dipolar anisotropy. This triggers a critical crossover at $T_c'$. Our study shows the emergence of a complex interplay of short- and long-range interactions below $T_c'$ as CGT approaches the 2D limit.

Amongst the popular two-dimensional (2D) van der Waal (vdW) materials[1-4] $Cr_2Ge_2Te_6$ (CGT) has attracted special attention as it is a 2D ferromagnetic (FM) - semiconductor (Curie Temperature $T_c$ ~ 67 K with a bandgap ~ 0.4 eV)[5] which is potentially useful for spintronic device applications[6-12]. These 2D-vdW materials offer controllability of their magnetic properties via, diverse parameters [2,7,13-18]. In CGT, strong Cr-Te-Cr super-exchange via a near $90^0$ bond angle[19,20] results in robust 2D - FM order. In general, dimensionality reduction can trigger, loss of inversion symmetry, modification of spin-orbit coupling (SOC)[21] which affects magnetic[12] and transport properties[22-24] and changes in band structure[25,26]. Critical phenomena exhibit characteristic scaling of data near a critical transition point[27], leading to the identification of universality classes. The universality class, determined solely by dimensionality, symmetry, and interactions in the system[28,29], allows for a deeper understanding of interactions driving phase transitions in condensed matter systems which is independent of the system's microscopic details. Therefore, exploring the nature of magnetic transition as a function of thickness in vdW materials is important. Magnetism of CGT has complex dimensional dependence: in bulk CGT, the magnetic anisotropy[1,30] is out of plane (OOP) i.e. along the normal to the crystallographic ab plane, however theoretical results on magnetic anisotropy in few-layer or monolayer (ML) CGT are conflicting, viz., one study suggests ML-CGT has OOP[31] anisotropy while another study suggests in-plane (IP)[32] anisotropy. Recent experimental studies in bulk CGT single crystals show the critical scaling exponents of magnetization (*M*) near $T_c$ conform to the 2D Ising-universality class[33,34] while another study suggests adherence to tricritical mean field model[35]. Some theoretical models[36] also treat bulk CGT as a 2D Heisenberg FM. However, the dimensional dependence of critical scaling features has not been systematically explored. In this work, using Modified Arrott Plot (MAP) method and critical scaling analysis,[29,37,38] we investigate the thickness (dimensional) dependence of $T_c$ and the critical scaling exponents in CGT. Bulk single crystal of CGT shows a 2D Ising-like paramagnetic (PM) to FM phase transition with OOP anisotropy at $T_c$ = 67 K. Few-layer (or low dimensional) CGT flake ensemble dispersed on hBN/SiO$_2$/Si substrate exhibits the same phase transition at 67 K. Additionally we also find a second critical transition at a lower temperature $T_c'$ ~ 14-15 K whose critical exponents are completely different from those at 67 K, and do not conform to the standard universality classes. Our DFT-based electronic structure calculations reveal small yet important structural changes in CGT at reduced dimensionality, which affect the spin orbit coupling (SOC). Our calculations suggest that in low dimensional CGT, a competition emerges between magneto-crystalline anisotropy energy (MAE) favouring OOP anisotropy and dipolar anisotropy energy (DAE) favoring IP orientation, and this results in a complex magnetic configuration. This also suggests the emergence of competing long and short-range interactions in the system with lowering dimensions, all of which, we believe, triggers a new critical crossover at $T_c'$.

Single crystals of CGT were grown by self-flux technique[39] (see details of experimental methods in SI-1, with X-ray Diffraction (XRD) characterization and Energy Dispersive X-ray (EDX) elemental map of the CGT crystal in SI-2). Crystals were exfoliated in Ar atmosphere in a glove box down to ~ 2 nm thickness. Fig.1(a) shows the similarity of the measured Raman spectral modes for a bulk single crystal and a 2 nm thick CGT flake. The spectra show two distinct peaks at 116 and 136 cm$^{-1}$ which correspond to the $E_g^2$ and $A_g^1$ Raman modes respectively in CGT[40,41]. Transmission Electron Microscopy (TEM) reveals the high degree of crystalline order of a few layered CGT flake (see TEM image of atomic planes in CGT in Fig. 1(b) and Fig. S2(e)). Fig. 1(c) shows a Z-contrast intensity profile along the red-marked region in Fig. 1(b). The atomic ordering profile (Fig.1(c)) agrees with CGT's simulated atomic configuration (Fig. 1(a) inset). The hexagonal symmetry of the diffraction spots in the Selected Area Electron Diffraction (SAED) image confirms the crystalline order persisting in the few-layer CGT (Fig. 1(d)) (see SI-2 for details). Carefully chosen few layered flakes were transferred onto insulating and inert single layer hBN/SiO$_2$/Si substrate (Graphene supermarket, see Fig. S2(h) in SI-2 for characterization). Fig. 1(e) shows the Atomic Force Microscopy (AFM) map of the transferred flakes recorded across multiple regions of area 10 μm × 3 μm. The uniformity of thickness (~2-3 nm) of the flakes in the ensemble is evident from the adjoining colour bar, confirming almost uniform 2D nature of the transferred flakes (see SI-3 for details). The typical density of the CGT flake ensemble (Fig.1(e)) is ~ 1 to 2 flakes/μm$^2$. In bulk CGT crystal, the magnetization (*M*) versus temperature (*T*) (SQUID magnetometer, Cryogenics UK) in an applied magnetic field (*H*) of 200 Oe, shows a PM to a FM transition (Fig. 2(a)) in both $H \parallel ab$ and $H \perp ab$ orientations. Inverse susceptibility ($\chi^{-1}$) versus $T$ derived from $M(T)$ data for $H \parallel ab$ (see inset of Fig. 2(a)) shows the typical Curie-Weiss nature. The intercept gives $T_c = 64.28$ K and an effective magnetic moment ($\mu_{eff}$) 3.18 $\mu_B$ per Cr atom, consistent with prior finding of orbital quenched magnetic moment of 3.87 $\mu_B$ of $Cr^{3+}$ in CGT[33]. The inset of Fig. 2(b) shows five-quadrant $M$ versus $H$ at $T = 4$ K for $H \parallel ab$ and $H \perp ab$. For $H \perp ab$, the $M(H)$ loop has a square shape with a saturation field $H_s^{\perp ab} \sim 3$ kOe and low coercivity $H_C \sim 240$ Oe, while for $H \parallel ab$, $M(H)$ has a skewed shape and does not saturate up to 40 kOe[42]. This suggests that the CGT single crystal has strong perpendicular magnetic anisotropy ($K_u$) i.e. the easy axis of $M$ is normal to the ab plane [1,42]. Linear extrapolation of the $M(H)$ data for $H \parallel ab$ gives $H_s^{\parallel ab} \sim 78$ kOe (see SI-4), using which we estimate the effective $K_u = 4.75 \times 10^5$ J.m$^{-3}$ ($\sim 0.6$ meV/f.u, which is close to the reported values[42,43] and also matches with our DFT results discussed later). Isothermal $M(H)$ for $H \perp ab$ were measured at different $T$'s (in intervals of 1 K) around $T_c$ for critical scaling analysis (see Fig. S5 in SI-5). In critical region ($T \to T_c$), we analyse $M(H)$ isotherms using the MAP method based on magnetic equation of state[38,44-46] $M(H,t) = t^\beta f_\pm \left(\frac{H}{t^{\beta+\gamma}}\right)$ (see details in SI-6), where $t = \frac{T-T_c}{T_c}$ and $f_\pm$ are two distinct functional forms of the scaled isothermal $M(H)$ curves at $T$ above (+) and below (-) $T_c$. In Fig. 2(b), we see that all the scaled magnetization $\left(\frac{M}{t^\beta}\right)$ versus scaled field $\left(\frac{H}{t^{\beta+\gamma}}\right)$ curves collapse onto two distinct family of curves, one above $T_c$ and another below $T_c$ for $\beta = 0.14$, $\gamma = 1.45$, and $T_c = 67$ K, thereby confirming the presence of a critical phase transition[38,44-46] close to 67 K. In Fig.2(c) the $M^{\frac{1}{\beta}}$ versus $\left(\frac{H}{M}\right)^{\frac{1}{\gamma}}$ shown using $\beta$ and $\gamma$ of Fig. 2(b) yields a set of parallel straight lines with the isotherm at 67 K passing through the origin, marking $T_c$. The $y$ and $x$ axis intercepts in Fig.2(c) give the $T$ dependent spontaneous magnetization $M_s$ (left axis Fig.2(d) for $T < T_c$) and inverse susceptibility ($\chi^{-1}$) (right axis Fig.2(d) for $T > T_c$) versus $T$, respectively. In Fig.2(d), the solid lines are fits to $M_s$ and $\chi^{-1}$ using well known power law fits[45]: $M_s(T) \propto t^\beta$ and $\chi^{-1}(T) \propto t^\gamma$, giving $\beta = 0.12 \pm 0.01$ and $\gamma = 1.21 \pm 0.34$, which are consistent with the estimates of Fig.2(b) and Fig. 2(c) (see supplementary section SI-7 for additional self-consistency checks of the critical exponents). Using $\beta = 0.14$, $\gamma = 1.45$ and the Widom's scaling relation[47] : $\delta = 1 + \frac{\gamma}{\beta}$ we find $\delta = 11.35(7)$. Our critical exponents for bulk CGT obtained are summarized in table I and they are not only consistent with earlier reports [33,34] (see table I) but are in good agreement with the 2D Ising universality class ($\beta = 0.125$, $\gamma = 1.75$).

Next, we measure the $M(T)$ and $M(H)$ of an ensemble of 2-3 nm thick CGT flakes (Fig.1(e)). Note that all the data shown here is after subtracting the diamagnetic background of the bare hBN/SiO$_2$/Si substrate (5 mm $\times$ 3 mm $\times$ 0.5 mm) and straw holder. We could get reliable, low-noise data for our few-layer CGT flakes only for $M(T)$ measured in the $H \perp$ ab orientation. $M(T)$ for the few-layered CGT flake ensemble (Fig. 3(a)) shows PM to FM transition with a $T_c \sim 66.4$ K deduced from the intercept of linear $\chi^{-1}$ versus $T$ plot (see Fig.3(a) inset), similar to the bulk CGT. For our few-layered CGT flake ensemble, Fig. 3(b) shows fit to the $M(T)$ data (FC 600 Oe) near the PM-FM transition with $M(T) \sim t^\beta$ which gives $\beta = 0.16 \pm 0.02$ and an extrapolated $T_c$ estimated to be $66.31 \pm 0.09$ K. Similar to Fig. 2(b) for bulk CGT, for our few-layered CGT flake ensemble, Fig 3(c) shows the scaled magnetization $\left(\frac{M}{t^\beta}\right)$ versus scaled field $\left(\frac{H}{t^{\beta+\gamma}}\right)$ curves collapse onto two distinct family of curves, one above $T_c$ and another below it, with $\beta = 0.16$, $\gamma = 1.44$ with an estimated $T_c = 65$ K. A log-log plot of the $M(H)$ data at 70 K (Fig. 3(d)) fits to $M(H) \sim AH^{1/\delta}$ with $\delta = 10.17$. We find that the critical exponents corresponding to PM-FM transition for few layered flakes at $T_c$ ($\beta = 0.16$, $\gamma = 1.44$, and $\delta = 10.17$) are close to the critical exponent values of the phase transition at $T_c$ in bulk CGT, viz., $\beta = 0.14$, $\gamma = 1.45$, $\delta = 11.379$. Thus, we find that the transition at $T_c$ in both bulk and the few layered CGT samples, both obey the 2D Ising universality class.

An additional feature to note is that in addition to the rise in $M$ near the PM-FM transition at $T_c$, there is a notable downturn in $M(T)$ near 20 K for the CGT flakes. A similar feature can be seen in recent $M(T)$ data of 14.4 nm thick epitaxial CGT film[48], although its origin has not been analyzed or discussed. Our studies reveal that the downturn in $M(T)$ at 20 K is robust and does not change with moderate $H$ variation (600 Oe to 1400 Oe). Fig. 3(e) shows $M(H)$ for different $T$ around this regime for the few-layer CGT flake ensemble. A Brillouin function fit[49] on $M(H)$ data at 2.2 K (see inset of Fig. 3(e)) gives $\mu_{eff} = 3.44\ \mu_B$ which is close to the bulk value of $3.18\ \mu_B$. Following MAP analysis, using $\beta = 0.15$ and $\gamma = 0.20$ a plot of $M^{1/\beta}$ versus $\left(\frac{H}{M}\right)^{1/\gamma}$ (Fig. 3(f)) gives a set of parallel straight lines with intercepts on either the $x$-axis (for $T > T_c$) or the $y$-axis (for $T < T_c$). The isotherm for 14.2 K passes through the origin, thereby identifying a new low $T$ critical transition temperature ($T_c'$) for the CGT flake ensemble. The $y$ and $x$-axis intercepts in MAP give $M_s(T)$ and $\chi^{-1}(T)$ for $T < T_c'$ and $T > T_c'$, respectively (see Fig. 3(g)). Fits with, $M_s(T) \sim t^\beta$ (for $T \leq 14$ K) gives $\beta = 0.21 \pm 0.01$ and $T_c' = (14.29 \pm 0.09)$ K and $\chi^{-1}(T) \sim t^\gamma$ (for $T > 15$ K) gives $\gamma = 0.18 \pm 0.01$ and $T_c' = (12.84 \pm 0.61)$ K. These values of $\beta$ and $\gamma$ are consistent with those in Fig. 3(f). Using $\beta = 0.15$ and $\gamma = 0.20$, Fig. 3(h) shows a bifurcation of the scaled $M(H)$ isotherms onto two distinct family of curves which suggests the presence of a critical transition at $T_c' = (14.0 \pm 1.0)$ K. The low $T$ critical transition window is narrowed down further (Fig. 3(i)) by noting a linear $M(H)$ trend in a log-log plot at 14.2 K, which suggests $\delta = 2.20$ (as $M = AH^{1/\delta}$ near the critical point). This value of $\delta$ agrees well with the value of 2.33 obtained from Widom's scaling relation (discussed earlier). The above analysis clearly shows the emergence of a new critical regime (see summary in Table I) at $T_c' \sim 14$ K in the few-layer flakes (2 to 3 nm thick) ensemble of CGT with a unique set of critical exponents distinct from those in the bulk crystal. We find none of the critical exponents for the transition at $T_c'$ resemble those for any known universality classes (see values mentioned in Fig. S6 in SI-6). In Figs. 4(a) and 4(b), we analyse the $M_s(T)$ data for $H \perp ab$ for few-layered flakes of CGT and bulk CGT at $T \ll T_c'$ and $T \ll T_c$, respectively. At low $T$ $\left(\frac{T}{T_c\ or\ T_c'} < 0.4\right)$, $M_s(T)$ fits well with the Bloch spin wave[50] model with a gapped ($\Delta$) magnon spectrum $\left(\sim T^{\frac{3}{2}} f_{\frac{3}{2}}\left(\frac{\Delta}{k_B T}\right)\right)$ where $f_{\frac{3}{2}}(y)$ is the Bose-Einstein integral function (see SI-9 for detailed discussions). From the fit, we get an anisotropy-induced gap $\Delta' = (0.14 \pm 0.03)$ meV in the magnon spectra for few-layered flake ensemble compared to $\Delta = (1.38 \pm 0.27)$ meV for bulk CGT. With reduced dimensionality the reduction in the gap of the magnon spectra strongly suggests a weakening of magnetic anisotropy at low dimension. It is not surprising that the ratio of the magnon gaps ($\Delta'/\Delta$) is similar to that of $T_c'/T_c \sim 0.2$.

**Table I**: Comparison of critical parameters for bulk CGT crystal and CGT flake ensemble

| Sample | $\mu_{eff}$ | $\beta$ | $\beta$ (Ref.[33,34]) | $\gamma$ | $\gamma$ (Ref.[33,34]) | $\delta$ | $\delta$ (Ref.[33,34]) | $T_c$ | $T'_c$ |
|---|---|---|---|---|---|---|---|---|---|
| CGT bulk crystal | $3.18\mu_B$ per Cr | 0.14 (Fig.2(b)) <br> 0.12 (Fig.2(d)) | 0.17 to 0.2 | 1.45 (Fig.2(b)) <br> 1.21 (Fig.2(d)) | 1.28 to 1.75 | 11.37 (Fig. S7(a)) <br> 11.35 (Widom) | 10.87 to 7.96 | 64.28 K (Fig.2(a)) <br> 63.68 K - 66.59 K (Fig.2(d)) <br> 67 K (Fig.2(b), (c)) | -- |
| CGT flake ensemble | $3.44\mu_B$ per Cr | 0.15 (Fig.3(f)) <br> 0.21 (Fig.3(g)) | -- | 0.20 (Fig.3(f)) <br> 0.18 (Fig.3(g)) | -- | 2.20 (Fig.3(i)) <br> 2.33 (Widom) | -- | 66.42 K (Fig.3(a)) | 14.29 K - 12.84 K (Fig.3(g)) <br> 14.80 K (Fig.3(h)) |

To understand the layer dependence of magnetic properties in CGT, we performed DFT-based first-principles electronic structure calculations (details in SI-8(A)). We optimized the lattice structure and atomic positions for bulk, bilayer, and monolayer CGT. Minor yet consequential changes in the lattice structure were found at reduced dimensions. The structural details are given in Table II. Notably, the in-plane lattice constant shrinks by 0.01 Å from bulk to bilayer and ~ 0.02 Å from bulk to monolayer. The nearest neighbor (NN) Cr-Cr distances decrease marginally in the bilayer and the monolayer compared to the bulk. The Cr-Te-Cr bond angles also decrease marginally. There are small distortions in the Te anti-prisms around the Cr atoms in the bulk so that all NN Te-Te distances in an atomic sub-layer are not equal. The variations in the NN Te-Te distances in the bulk is ~ 0.07 Å, which reduces to 0.03 Å in the bilayer. And in monolayer, these variations disappear entirely so that all NN Te-Te distances are equal. The magnetic moment ($\mu$) comes largely from the Cr atoms, and it is 3 $\mu_B$/Cr in all three cases, in good agreement with experiments, suggesting $\mu$ is independent of the number of layers.

**Table II**: Structural details and anisotropy energies for bulk, bilayer and monolayer CGT

| System | a(Å) | b(Å) | Cr-Cr(Å) | Cr-Te(Å) | Te-Te(Å) | ∠Cr-Te-Cr | Interlayer Distance (Å) | MAE (meV/f.u.) | DAE (meV/f.u.) | $\epsilon_{ani}$ (meV/f.u.) |
|---|---|---|---|---|---|---|---|---|---|---|
| Bulk | 6.892 | 6.892 | 3.98 | 2.78 | 3.765 ± 0.035 | 91.29° | 3.38 | 0.3 | 0.095 | 0.395 |
| Bilayer | 6.882 | 6.882 | 3.97 | 2.78 | 3.755 ± 0.015 | 90.96° ± 0.12° | 3.41 | 0.143 | -0.022 | 0.121 |
| Monolayer | 6.875 | 6.875 | 3.97 | 2.79 | 3.75 | 90.72° | - | 0.066 | -0.021 | 0.045 |

We also calculate the MAE originating from SOC for bulk, bilayer, and monolayer CGT. MAE is the difference between the total energies of IP and OOP orientations of the $Cr^{3+}$ spins. A positive (negative) value of anisotropy energy indicates OOP (IP) anisotropy. Net anisotropy energy ($\epsilon_{ani}$) is the sum of MAE and DAE, the latter arising due to long-range magnetostatic interactions between the magnetic dipole moments on the $Cr^{3+}$ ions. DAE also depends on the orientation of the moments (see SI-8(B) for details of DAE calculation). Results for MAE, DAE, and $\epsilon_{ani}$ for each system are shown in Table II. The positive sign for MAE shows a preference for OOP orientation of the $Cr^{3+}$ moments from bulk to monolayer CGT. However, table II also shows that the magnitude of MAE decreases drastically (~75%) from the bulk to the monolayer. In CGT, the contributions from the highest atomic number Te atoms dominate the SOC strength. To understand the variation in the contributions from the Te atoms due to dimensional reduction, the total density of states (DOS) and its projections onto each atom (partial DOS) are shown in Fig. 4(c), (d), and (e). States near the valence band maximum have significant contributions from the Te *s* and *p* states in the bulk. However, in the bilayer and monolayer, these contributions are gradually suppressed. This suppression of contribution from Te states reduces the SOC, and hence MAE reduces significantly in low-dimensional CGT. The reduction of MAE should alter the spin gap in low dimensional CGT (compared to the bulk), as evidenced in Fig. 4(a).

In bulk, MAE and DAE cooperatively support OOP orientation of $Cr^{3+}$ moments. This is seen in Fig. 2 and is also consistent with earlier results[1]. However, we see DAE changing sign for bilayer and monolayer cases, favoring IP anisotropy. With reducing dimensionality MAE values reduce, but it still favours OOP orientation. Table II reveals this emergent competition between MAE and DAE in CGT, resulting in the decrease of $\epsilon_{ani}$ with reducing dimensionality. The above features are described by the Hamiltonian, $\mathcal{H} = \mathcal{H}_{int,z} + \mathcal{H}'_{xy}$, where $\mathcal{H}_{int,z} = -J_z \sum_{\langle ij \rangle} s_z^i s_z^j$ is the short-range NN (*i* and *j* are site index) Ising Hamiltonian favoring OOP FM ordering for $J_z > 0$. The $\mathcal{H}'_{xy} = (-J'_d \sum_{\langle ij \rangle} [s_x^i s_x^j + s_y^i s_y^j] + \mathcal{H}_{dip})$ term describes interactions between the in-plane spin components with $J'_d$ representing a dimension (*d*)-dependent exchange interaction and $\mathcal{H}_{dip}$ is the dipolar interaction term which is also *d*-dependent (representing the DAE contribution). In bulk CGT (*d* = 3) we consider, $J_z \gg J'_d$ and $\mathcal{H}_{int,z} \gg \mathcal{H}'_{xy}, \mathcal{H}_{dip}$. Thus $\mathcal{H}_{int,z}$ is predominantly responsible for OOP critical fluctuations leading to 2D Ising universality class at the critical transition $T_c$ in bulk CGT. Our DFT results show structural distortions increase with reduced dimensionality which decrease SOC and hence MAE, i.e., $\mathcal{H}_{int,z}$ weakens (Table II) [18,19,51] and concomitantly DAE increases. Theoretical studies based on exchange interaction models[52] of the type $\mathcal{H} = (\mathcal{H}_{int,z} + \mathcal{H}'_{xy})$, suggests the emergence of a new critical crossover regime associated with enhanced fluctuations along in-plane *x* and *y* directions (due to $\mathcal{H}'_{xy}$ contribution). A critical crossover separates two regimes with very different universality classes[53]. In our experiments, a similar critical crossover emerges at $T'_c$, below which emerges a unique critical scaling behavior. At low $T$ ($< T'_c$), low dimensional CGT enters a magnetically inhomogeneous configuration due to the competition between OOP ($\mathcal{H}_{int,z}$, MAE) and IP ($\mathcal{H}'_{xy}$, DAE) anisotropies. As net *M* contribution along the measuring direction ($H \perp$ ab) decreases, there is a turnaround at $T'_c$ (Fig. 3(a)). The strengthening of DAE suggests the emergence of long-range dipolar interactions which coexists along with short range interactions at low *d*. Due to vdW stacking of 2D layers, naively one may expect a preservation of the universality class in transforming from bulk to 2D - CGT layers. However, this expectation clearly breaks down, as we see the emergent new critical crossover regime at $T_c'$ in low dimension CGT. Here the combined presence of short- and long-range interactions in 2D-CGT results in critical scaling properties deviating significantly from known universality classes.

To summarize our findings, our experiments confirm that the PM - FM phase transition in bulk CGT at $T_c$ (~ 67 K) obeys the 2D Ising universality class. However, in the few-layered CGT flake, in addition to the usual 2D Ising like PM-FM transition at $T_c$, we report the presence of another critical transition at a much lower $T$ (viz., $T_c' \sim 14$ K). The transition at $T_c' \sim 14$ K is characterized by a new set of critical exponents (Fig. 3(f)-(i)) which are very different from the 2D Ising universality class. This change in the universality class below $T_c'$, suggests an alteration in the nature and range of interactions in few-layer thick CGT as compared to the bulk samples. From the gapped spin wave analysis of $M_s(T)$, we found out that the magnetic anisotropy reduces by an order of magnitude with thinning down of CGT down to a few-layers from the bulk (Fig. 4(a) and (b)). Our DFT based electronic structure calculations reveal that with reduced dimensionality (i.e., thickness of CGT), there are subtle structural reorganisation occurring in the system, which affects SOC which in turn modifies the MAE. Our calculations show that MAE decreases significantly at reduced CGT dimensionality (without changing its sign) while the DAE changes significantly compared to its value in bulk CGT. In few layer thick CGT the DAE changes sign, suggesting that the IP magnetization orientation at lower dimensions is favoured compared to the OOP orientation. It may be noted that these calculations are done effectively at $T = 0$ K, i.e., not considering any thermal fluctuation effects. While in bulk CGT, MAE and DAE both co-operatively favour OOP orientation of moments at $T_c$, however at lower dimensionality there is a preference for IP orientation of moments, resulting from the competition between MAE and DAE at $T_c'$. Thus reduced dimensionality in CGT gives rise to a competition between two anisotropy energy scales at a low $T$. This competition isn't strong near $T_c$, as DAE is weak near $T_c$ because of thermal fluctuation of moments near this critical regime around $T_c$. Hence, just below $T_c$, the spins collectively order into a FM state obeying the 2D Ising universality class, which is predominantly governed by the short-range exchange interaction and MAE, in both bulk and few-layer samples. Therefore, both bulk and few layer CGT samples show identical 2D-Ising like universality class near $T_c$. However, with further lowering of $T$ below $T_c$, the effects of DAE strengthens as the thermal fluctuation effects weaken. Eventually at $T \leq T_c'$ (~ 14 K), there appears the competition between MAE and DAE (with sign change) (see Table II). Consequently, there is a crossover from short range interaction dominated regime near $T_c$ into a regime with both, short- and long-range interactions active below $T_c'$. In this low $T$-regime, in the few-layer thick CGT sample, there is a critical crossover from a 2D Ising like state into a regime characterized by a new universality class. This regime characterised by MAE and DAE (sign change) competition, could possibly be a magnetically inhomogeneous state with an admixture of OOP and IP orientation of moments. It is also possible that weak structural modifications (which affect MAE) may also become significant at low $T$ (well below $T_c$) in the low dimensional CGT. This may in turn be an additional factor aiding the competition between MAE and DAE playing up at $T_c'$ in few layer thick CGT. Furthermore below $T_c'$ as long-range dipolar interaction effects are active, hence in low dimensional CGT sample, local fluctuations may begin to affect the ordering at larger length scales. A study of such rich emergent features, new universality classes and similar exploration in other low dimensional 2D-vdW magnetic systems, calls for further detailed experimental and theoretical investigations.

**Acknowledgement:** The authors acknowledge Advanced Imaging Centre, IIT Kanpur for providing HR TEM facility. S. S. Banerjee thanks the financial support from DST-SUPRA and DST - AMT of GOI India programs as well as for research infrastructure and funding from IIT Kanpur. All computations have been done on the HPC cluster facility at HRI (https://www.hri.res.in/cluster/). Suprotim Saha acknowledges the Prime Minister's Research Fellows (PMRF) Scheme of the Ministry of Human Resource Development, Govt. of India, for funding support. SD acknowledges the financial support from DST-SERB (grant no. CRG/2021/004334) and TRC, IACS. BD is grateful to IACS for the fellowship.

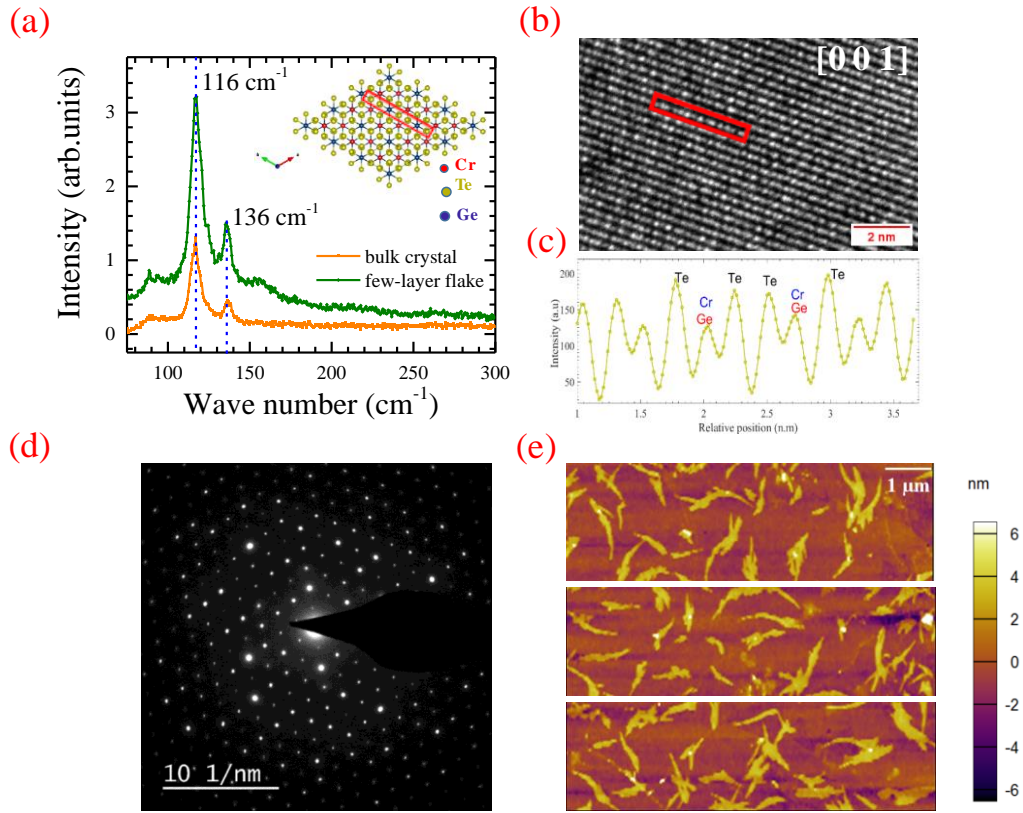

FIG. 1 (a) Raman spectrum of bulk single crystal of CGT and 2 nm thick CGT flake. (inset) Top view of simulated atomic model of CGT. (b) Atomic resolution HR-TEM image of a few-layered CGT flake, viewed from [001] direction. (c) Z-contrast intensity profile of the red-marked region in Fig. 1(b). Compare the profile to the similar red-marked region in Fig. 1(a) inset. (d) SAED spot pattern of a few-layer CGT flake. (e) AFM map at three different regions of area 10 μm × 3 μm showing multiple CGT flakes. Note the adjoining colour bar denoting thickness, indicating uniformity of thickness of the flakes (cf. text for details).

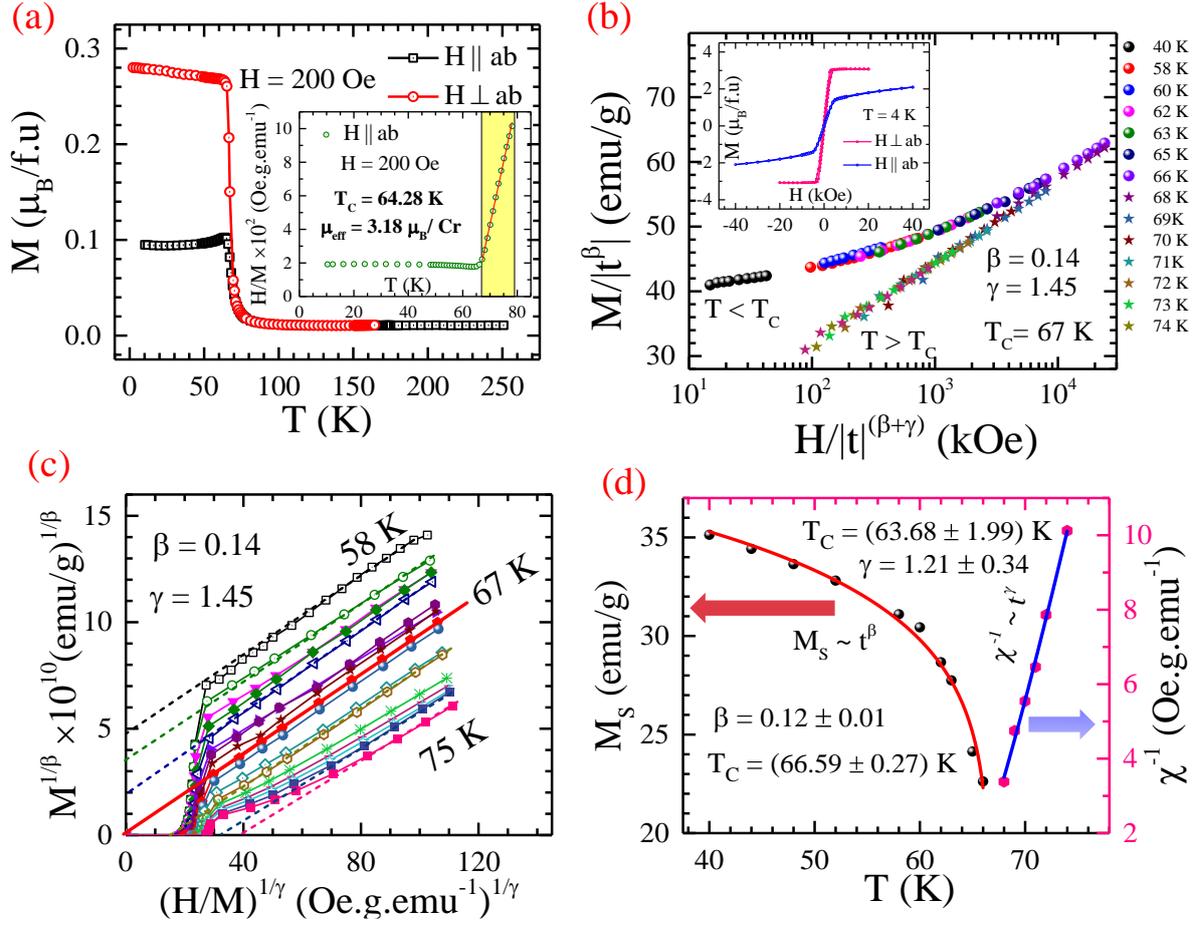

FIG. 2 (a) Field cooled $M(T)$ for bulk CGT for $H \parallel ab$ and $H \perp ab$ at $H = 200$ Oe. (inset) Curie-Weiss fit to $\chi^{-1}(T)$ for $H \parallel ab$. (b) Scaled magnetization versus scaled field in the critical regime bifurcating above and below $T_c = 67$ K. (inset) $M(H)$ for bulk CGT at 4 K for $H \parallel ab$ and $H \perp ab$. (c) MAP in the critical regime with $\beta = 0.14$ and $\gamma = 1.45$. (d) $M_s(T)$ (left axis) and $\chi^{-1}(T)$ (right axis) deduced from intercepts on $y$-axis ($T<T_c$) and $x$-axis ($T>T_c$) ) respectively of Fig. 1(c) along with fits through data (cf. text for details).

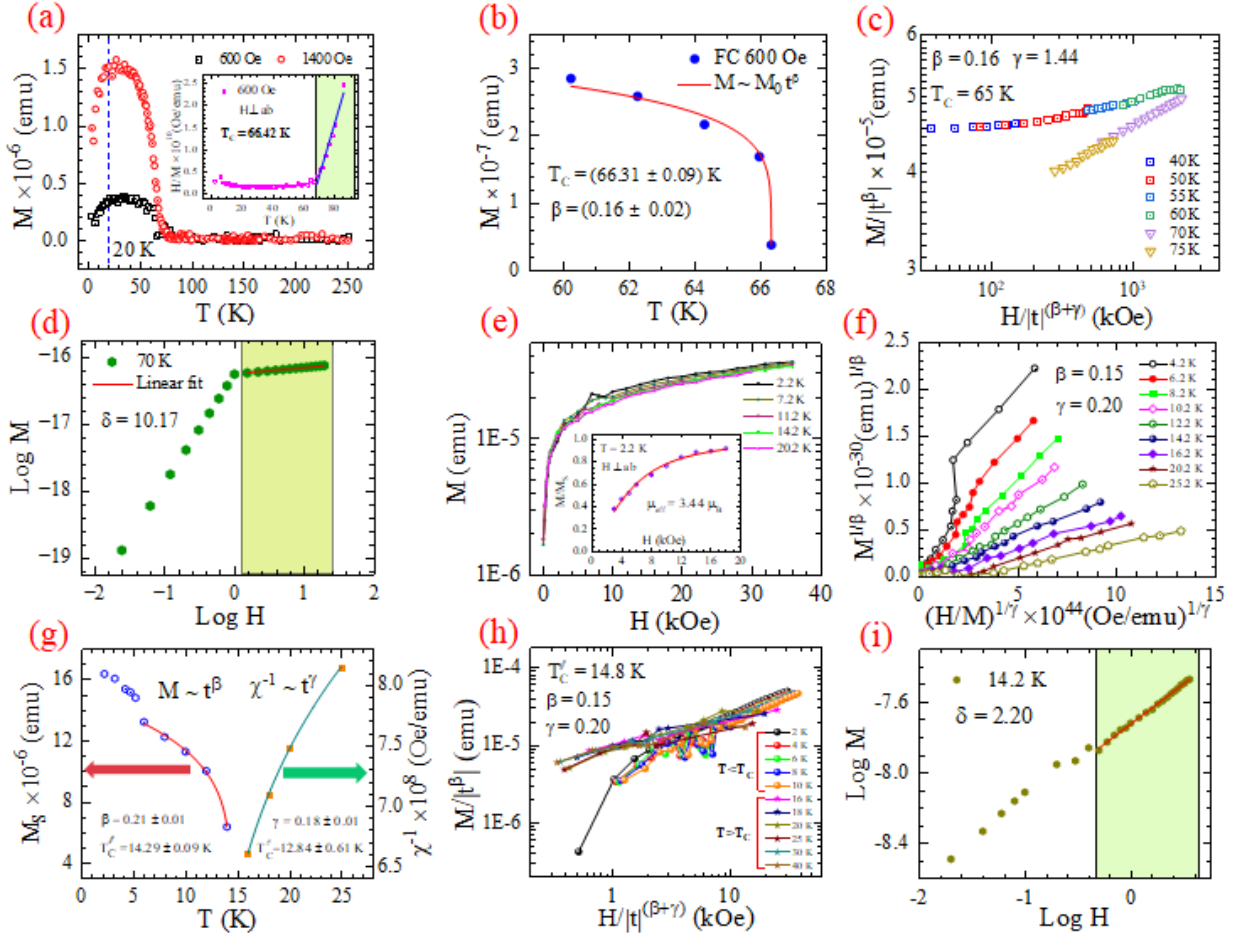

FIG. 3 (a) $M(T)$ for few layer flake-ensemble of CGT for out of plane field 600 Oe and 1400 Oe. (inset) Curie-Weiss fit to $\chi^{-1}(T)$ for $H = 600$ Oe. (b) $M(T)$ for 600 Oe fitted with power law near $T_c$. (c) Scaled magnetization versus scaled field showing bifurcation above and below $T_c$ with $\beta = 0.16$, $\gamma = 1.44$ and $T_c = 65$ K. (d) $M(H)$ isotherm in log-log scale at 70 K to determine $\delta$. (e) Background-corrected representative $M(H)$ data of CGT flake-ensemble from $T = 2.2$ K to 20.2 K. (inset) Brillouin fit to $M(H)$ at 2.2 K. (f) MAP for isotherms in the critical regime ($T \sim T_c'$) with $\beta = 0.15$ and $\gamma = 0.20$. (g) $M_S(T)$ (left axis) and $\chi^{-1}(T)$ (right axis) with respective power law fits in $T < 14$ K and $T > 15$ K respectively. (h) Scaled magnetization versus scaled field showing bifurcation above and below $T_c' = 14.8$ K. (i) $M(H)$ isotherm in log-log scale at 14.2 K to determine $\delta$ (cf. text for details).

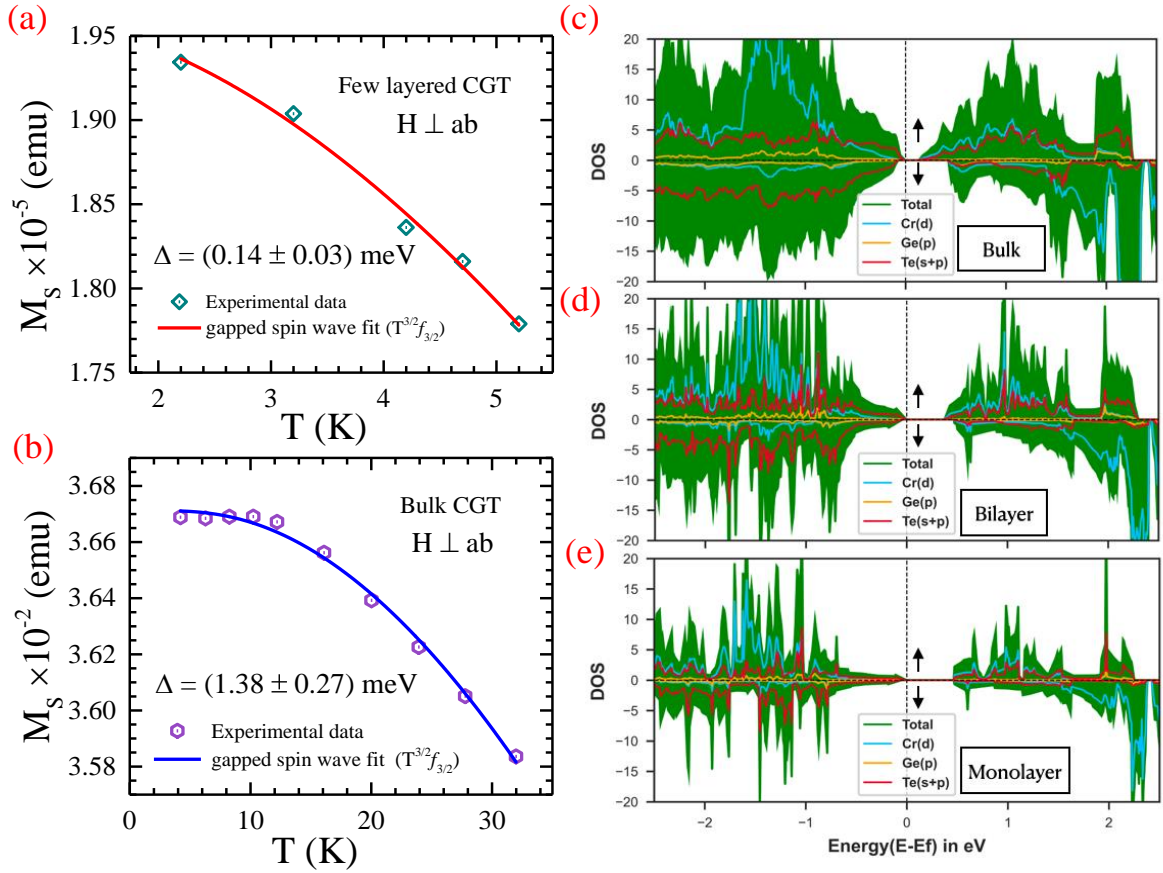

FIG. 4 (a) $M_s(T)$ for few-layered CGT flakes at $T < 5.2$ K for $H \perp$ ab with fit to gapped spin wave dispersion. (b) $M_s(T)$ for CGT bulk crystal at $T < 32$ K for $H \perp$ ab with fit to gapped spin wave dispersion. (c)-(e) Projected DOS of CGT in (c) bulk, (d) bilayer and (e) monolayer form (cf. text for details).

**Supplementary Information**

**SI-1: Details of experimental methods**

Sample growth and characterization of bulk crystal

Single crystals of CGT were grown by self-flux technique using pure constituent elements Cr, Ge and Te powders mixed with molar ratio 1:2:6 in an evacuated quartz tube. The mixture was heated from room temperature to 1323 K at a heating rate of 100 K/hour and kept at that temperature for 24 hours followed by a slow cooling. Crystals thus obtained are typically mm-sized, exfoliable and had a shiny plate-like appearance. X ray Diffrcation (XRD) was performed on a crystal of size 2.5 mm x 2.3 mm x 0.24 mm using a Panalytical X-ray diffractometer with Cu-K$\alpha$ radiation ($\lambda$ = 1.5406 Å) at room temperature in the 2$\theta$ range of (10º–90º) with a precision of 0.05º. CGT single crystals were characterized by pre-calibrated Raman spectroscopy (Princeton Instruments Acton Spectra Pro 2500i ) with a 785 nm LASER incident from (001) direction in ambient condition. A laser power of 2 mW with a spot-size of ~10 μm was used with a grating of 1200 rulings per mm. Proper care was taken to limit the LASER power to ensure no heating/burning effect on the crystals. Raman spectra taken at different regions of the bulk single crystal and from various flakes of similar thickness (2-3 nm) yield identical results. Note that the single crystals used for Raman characterization are exposed to air, which has an effect of suppressing the peaks at higher wave numbers ($E_g^3, E_g^4, A_g^2$)[1]. Also note the enhancement in Raman intensity for the nanoflakes compared to that of the bulk[1]. In order to confirm homogeneity in chemical composition, Energy dispersive X ray spectroscopy (EDX) is measured in a JEOL JSM-6010LA Scanning Electron Microscope (SEM) using a 20 kV electron beam. High resolution Transmission electron microscopy (HR-TEM) and Selected Area Electron Diffraction (SAED) of the CGT single crystals were performed at ambient temperature using a FEI Titan G2 60 -300 with a 300 kV electron beam from (001) direction for structural investigation. CGT single crystals were exfoliated in a glove-box maintained in inert Argon atmosphere and fine crystal pieces were floated in xylene and this diluted solution containing CGT flakes was subsequently drop-cast on a carbon coated copper grid for TEM and SAED. The interplanar spacings ($d_{hkl}$) were calculated by measuring the distances between the central diffraction spot and the diffracted bright spots using ImageJ software[2] and individual bright spots were indexed to (h k l) planes using the calculated interplanar spacings by Vesta [3] as reference.

Preparation of few layered samples and thickness characterization

Crystals were exfoliated by adhesive tape in a glove-box maintained in inert Argon atmosphere and thin flakes carefully examined by optical contrast were transferred on single layer hBN/SiO$_2$ (285 nm)/Si (500 μm) substrate (purchased from Graphene supermarket) using PDMS stamp for magnetic measurement. Thickness of the transferred flakes were determined by ac-tapping mode Atomic Force Microscopy (AFM) measurement using a Al(100) coated Si probe-tip with typical frequency ~70 kHz and spring constant 2 N/m in an AC240TS-R3 model Asylum Research AFM. Considering the softness of the flakes, contact mode measurement was avoided. Note that AFM measurements often have a tendency to overestimate the thickness of flakes owing to the trapping of air and moisture at the flake-substrate interface.

Magnetic measurements:

All magnetic measurements were performed in a commercial S700X Superconducting Quantum Interference Device (SQUID) magnetometer (CRYOGENIC, UK) allowing a dc sensitivity of $10^{-8}$ emu in terms of measurement of magnetic moment. Highly homogeneous magnetic field from a superconducting magnet ($T_c \sim 9$ K) is used. The interior of the cryostat (and the sample space) is screened from external magnetic field (including terrestrial magnetic field) by a mu-metal shield. Samples (both bulk and few-layered flakes) were loaded in a non-magnetic capsule inside a glove box with the desired crystallographic orientation with respect to the magnetic field ($H \parallel$ ab or $H \perp$ ab) and mounted in the cryochamber with as little delay as possible (typically 30-60 s). Each measurement of magnetic moment is determined by averaging three scans in same thermomagnetic condition. The non-hysteretic diamagnetic contribution in *M-H* (linear trend in second and fourth quadrant) due to the hBN/SiO$_2$/Si substrate (and straw holder) was subtracted from the raw data and this background corrected data was eventually used for analysis. For both bulk and few-layered cases, sample was warmed to room temperature and demagnetized, between measuring each *M(H)* isotherm.

## SI-2: Structural information and chemical composition

Cr$_2$Ge$_2$Te$_6$ is known to crystallize in a rhombohedral $R\bar{3}$ structure [4]. Bulk Cr$_2$Ge$_2$Te$_6$ forms layered crystals with an interlayer vdW gap of 6.81 Å (Fig. S2(a)). With a weak vdW force acting between adjacent layers, the material is easily cleavable down to monolayer. A view of the monolayer along [0 0 1] shows (inset of Fig. 1(a) in MS) each Cr atom is octahedrally coordinated to six Te atoms, also notice the hexagonal sublattice formed by the Cr atoms. A typical single crystal of dimensions 2.5 mm x 2.3 mm x 0.24 mm is shown in the inset of Fig. S2(b). The XRD 2θ scan (Fig. S2(b)) shows (0 0 *l*) peaks, implying stacking of layer takes place along (0 0 1) direction in CGT single crystal. Analysis of our XRD pattern gives the lattice parameter *c* = 20.56(7) Å, in good agreement with earlier report[4,5]. The EDX elemental map of Cr, Ge and Te in a 15 μm × 12 μm area (Fig. S2(c)) shows the chemical homogeneity in our CGT sample. The elemental stoichiometry and chemical homogeneity is measured from individual point EDX measurement (Fig. S2(d)), which yields an atomic percentage ratio of 19.11

: 17.71 : 63.18, for Cr, Ge, and Te respectively, in good agreement with expected stoichiometry 2 : 2 : 6.

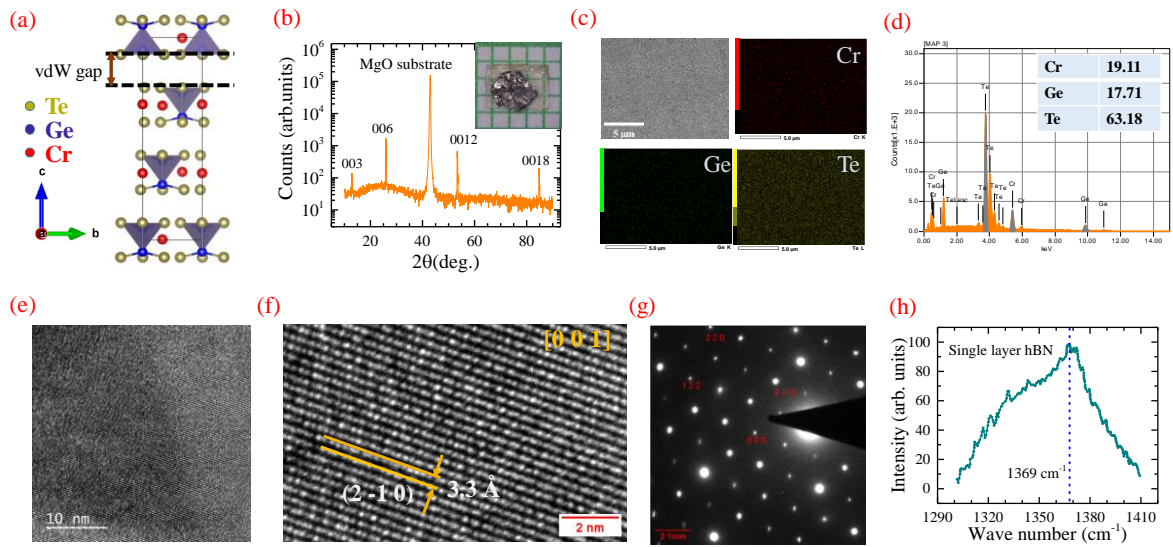

FIG. S2(a) Simulated crystal structure of CGT, viewed from a-axis, black rectangle denoting the unit cell. Note the vdW gap between adjacent planes. (b) XRD 2$\theta$ scan on CGT single crystal, (inset) a CGT single crystal on MgO substrate. (c) SEM-EDX elemental map of a CGT single crystal (d) Individual point EDX spectrum of CGT confirming elemental stoichiometry 2:2:6 (e) Real space TEM image of a few layered CGT flake, viewed from [001] direction (f) Atomic resolution HR-TEM image of few-layered CGT crystal viewed from [001] axis. Yellow parallel lines mark a set of planes (2 -1 0). (g) Corresponding SAED spot pattern, several individual spots are marked with respective Miller indices, including (2 -1 0). (h) Raman spectroscopy measurement of single layer hBN.

HRTEM with a 300 kV e-beam from [0 0 1] direction on a few-layered CGT crystal shows perfect crystalline nature with no hint of dislocations and grain boundaries within the imaged area (Fig. S2(e) and S2(f)). Six-fold symmetry in the SAED dot pattern (Fig. S2(g) and Fig. 1(d) in MS) taken along [0 0 1] zone axis confirms the single-crystalline nature, indicating absence of grain boundaries or rotational boundaries. The interplanar spacing for the yellow-marked parallel set of planes in Fig. S2(f) was measured from the HRTEM image to be 3.3 Å, which is identified as the (2 -1 0) plane for CGT[6] from calculated powder XRD pattern. We marked the corresponding diffraction spot in SAED image (Fig. S2(g)). Likewise, other lattice planes are identified by their interlayer spacings and diffraction spots corresponding to those spatial frequencies are labelled in the SAED image. Note the bright diffraction spots conform to the rhombohedral existence law $-h + k + l = 3n$ [4]

Single layer hBN (on $SiO_2$/Si) substrate was characterized by its Raman signature. A typical Raman spectroscopy measurement (Fig. S2(h)) shows a peak at 1369 cm$^{-1}$, consistent with earlier report[7].

# SI-3: Distribution of CGT flakes on hBN/SiO$_2$/Si and characterization of thickness

AFM maps show dense population (~1-2 flakes/$\mu m^2$) of few-layered CGT flakes (Fig. 1(e) in MS). Thicknesses of individual flakes from different regions are inferred from line cuts, shown below (Fig. S3 (a)-(l)). Also note the step of height 0.8 nm (Fig. S3(f)) which is equal to the interlayer distance for CGT[8,9].

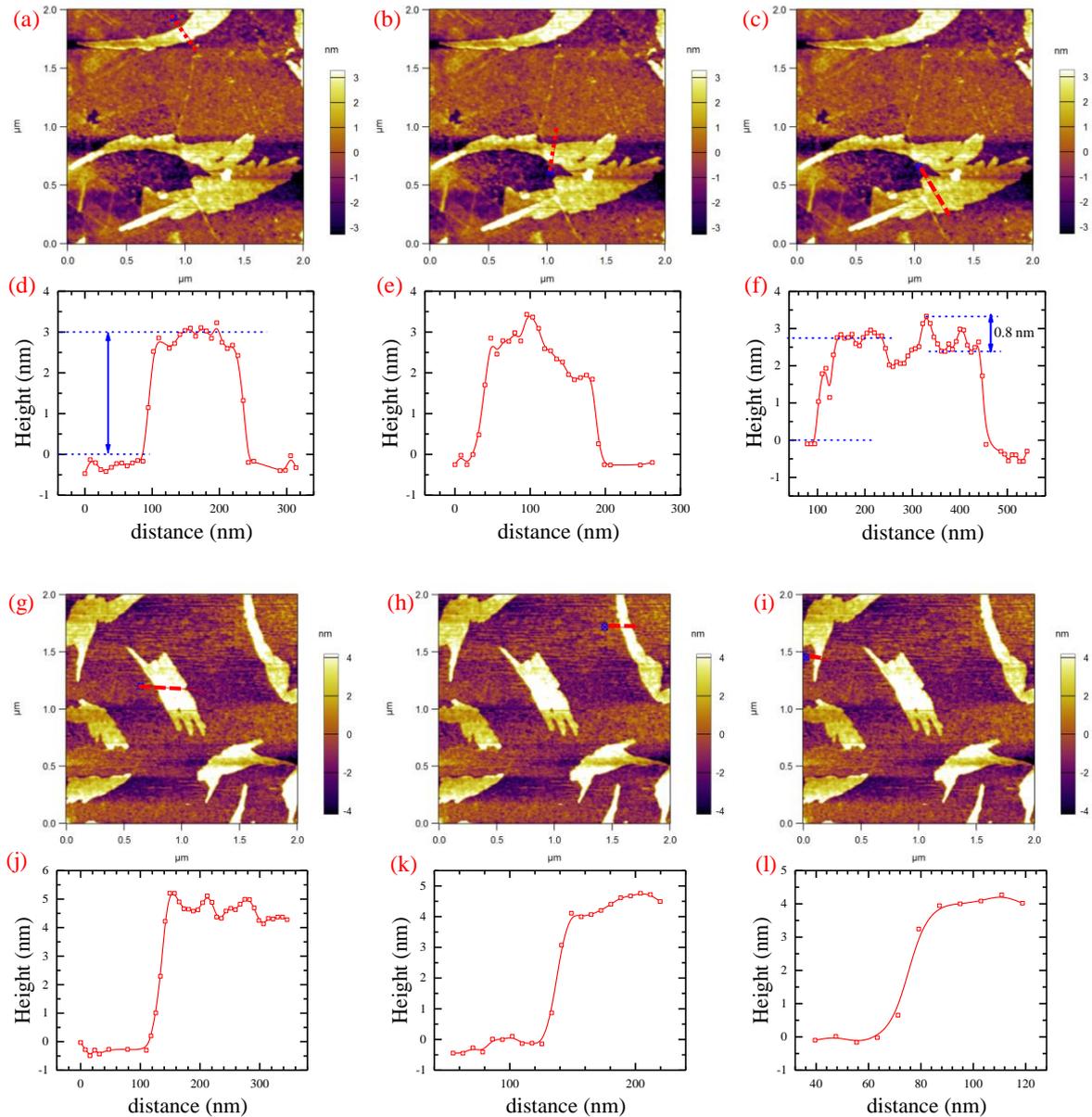

FIG. S3(a)-(c) AFM images of CGT flakes (dispersed on hBN/SiO$_2$/Si substrate) in 2 μm × 2 μm area. Note the adjoining colour bars, denoting thickness. (d)-(f) AFM height profile along the red dashed lines in (a)-(c) respectively. (g)-(i) AFM images of CGT flakes (dispersed on hBN/SiO$_2$/Si substrate) at a different location. (j)-(l) AFM height profile along the red dashed lines in (g)-(i) respectively.

## SI-4: Anisotropy in CGT and its experimental quantification

Our magnetic measurements on bulk single crystals of CGT show presence of significant anisotropy. $M(H)$ at 4 K (inset of Fig. 2(b) in MS) shows the out of plane orientation is the easy axis for Cr moments which is believed to the mediated by Cr-Te-Cr superexchange and single ion anisotropy[10,11]. We determine the saturation field for $H \parallel ab$ ($H_s^{\parallel ab}$) through linear extrapolation of the $M(H)$ curve (see dashed green lines in Fig. S4(a)) and calculate the magnetocrsytalline anisotropy energy $K_u$ from the formula $K_u = \frac{1}{2} M_s H_s^{\parallel ab}$ [12]. This estimation of $K_u \sim 4.75 \times 10^5$ Jm$^{-3}$ (or 0.6 meV/f.u) is consistent with earlier report[13] and our DFT results (cf. Table II in MS). $K_u$ is observed to be depending on $T$ (Fig. S4(b)) and we compare this $K_u(T)$ dependence with the $M_s(T)$ dependence (Fig. S4(c)) and further analyze in the light of Callen-Callen power law[14], which infers about the difference in $T$-dependence of anisotropy constant ($K_u(T)$) for uniaxial and cubic (or multiaxial) systems depending on crystal symmetry. Callen-Callen power law gives a relation between the $T$-dependence of $K_u$ and $M_s$

$$\frac{K_u(T)}{K_u(T=0)} = \left(\frac{M_s(T)}{M_s(T=0)}\right)^n \quad \text{------------------------(S1)}$$

where the exponent $n$ depends on the crystal symmetry and is expected to be 3 for uniaxial anisotropic system. However, for our case, the exponent is close to 2, differing from Callen-Callen prediction for uniaxial anisotropy. This deviation suggests that anisotropy in CGT has a complex origin, beyond the simplistic assumptions of Callen-Callen.

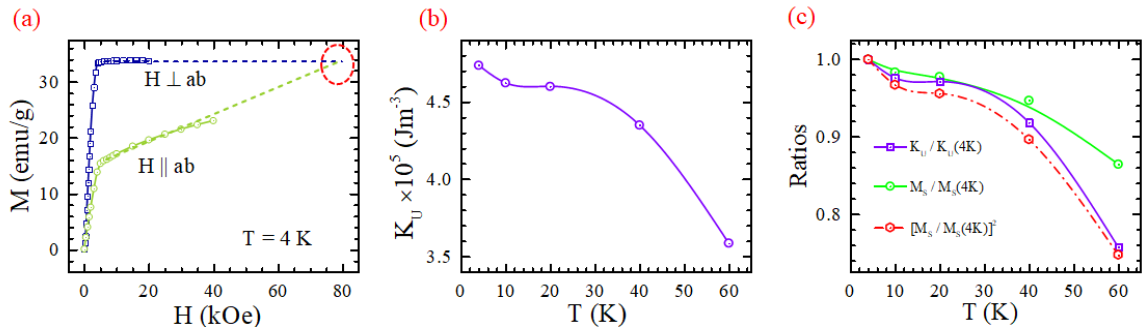

FIG. S4(a) $M(H)$ for bulk CGT at 4 K for $H \parallel ab$ and $H \perp ab$. The experimental data for both cases are linearly extended to find out saturation field for the hard plane ($H_s^{\parallel ab}$). (b) Experimental quantification of magnetic anisotropy $K_u(T)$; violet curve is a guide to eye. (c) The ratio $K_u(T)/K_u(4\ K)$ and $M_s(T)/M_s(4\ K)$ as a function of temperature (violet and green solid curves, respectively) plotted in comparison to $(Ms(T)/Ms(4\ K))^n$ for n = 2 (red dashed curve).

## SI-5: Magnetization isotherms for bulk CGT

$M(H)$ isotherms taken up to 30 kOe along the easy axis ($H \perp ab$) close to the critical regime of bulk CGT is shown in Fig.S5. Note that after measuring each $M(H)$ isotherm, the sample was warmed to room temperature and demagnetized before measuring the next $M(H)$ isotherm.

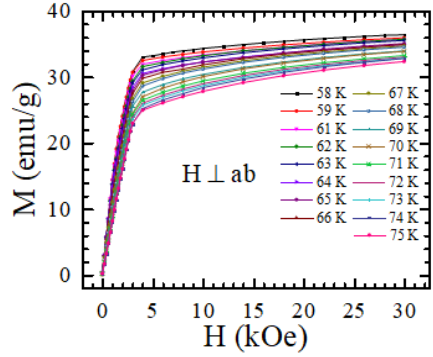

FIG. S5 Isothermal $M(H)$ curves for bulk CGT near the critical region (58 K to 75 K).

## SI-6: Methodology to evaluate critical exponents through Modified Arrott Plot (MAP)

It is known experimentally that close to a second order phase transition, experimental observables like $M_s$ and $\chi$ exhibit certain power law behaviours and each observable is assigned an exponent, given by the following equations[15] :

$$M_s = M_0(-t)^\beta \quad \text{for } t < 0 \quad \text{------------------ (S2)}$$

$$\chi^{-1}(T) = \chi_0^{-1} t^\gamma \quad \text{for } t > 0 \quad \text{------------------ (S3)}$$

$$M = AH^{1/\delta} \quad \text{for } t = 0 \quad \text{------------------ (S4)}$$

where $t = \dfrac{T - T_c}{T_c}$ is the reduced temperature and $\beta, \gamma, \delta$ are the critical exponents.

A proper choice of critical exponents $\beta$ and $\gamma$ gives a family of parallel straight lines[16] as $M(H)$ isotherms in the critical regime are plotted in the form $M^{1/\beta}$ vs $\left(\dfrac{H}{M}\right)^{1/\gamma}$. Arrott Plot for the Mean field model ($\beta = 0.5$ and $\gamma = 1.0$) using data shown in Fig. S5 is shown in Fig. S6(a). Even a cursory glance at Fig. S6(a) shows that no isotherm passes through the origin and clear non-linearity and downward curvature is observed for all isotherms, indicating mean field picture is not an appropriate description of the critical phenomenon in bulk CGT. In order to find out these exponents, we perform the iterative procedure detailed below[17]:

1. We plot MAPs using $\beta$ and $\gamma$ of several universality classes (Fig. S6(a)-(f)) and eventually choose 2D Ising model (Fig. S6(f)) as our most promising initial guess for the parameters $\beta$ and $\gamma$ because the family of curves are significantly linear for these choices of $\beta$ and $\gamma$ (note the clear downward curvature in Fig. S6(a)-(e)).

2. Using the MAP for 2D Ising model, we find out the resulting $M_s$ and $\chi^{-1}$ for all isotherms from the linear extrapolation of the curves on vertical and horizontal axes respectively.

3. $M_s(T)$ and $\chi^{-1}(T)$ data are fitted with eqn. (S2) and eqn. (S3) respectively, (while $T_c$ is kept a free parameter) to obtain a new set of $\beta$ and $\gamma$. These values of $\beta$ and $\gamma$ are used again to construct a new MAP.

4. The process is continued until $\beta$ and $\gamma$ from these two methods (i.e. MAP and fitting of $M_s(T)$, $\chi^{-1}(T)$) converge sufficiently close.

We emphasize that this iterative method is a general method and works equally well for any initial choice of $\beta$ and $\gamma$. However, for the method to converge quickly, it is advisable to start with a premeditated guess compatible with the critical phase transition under investigation.

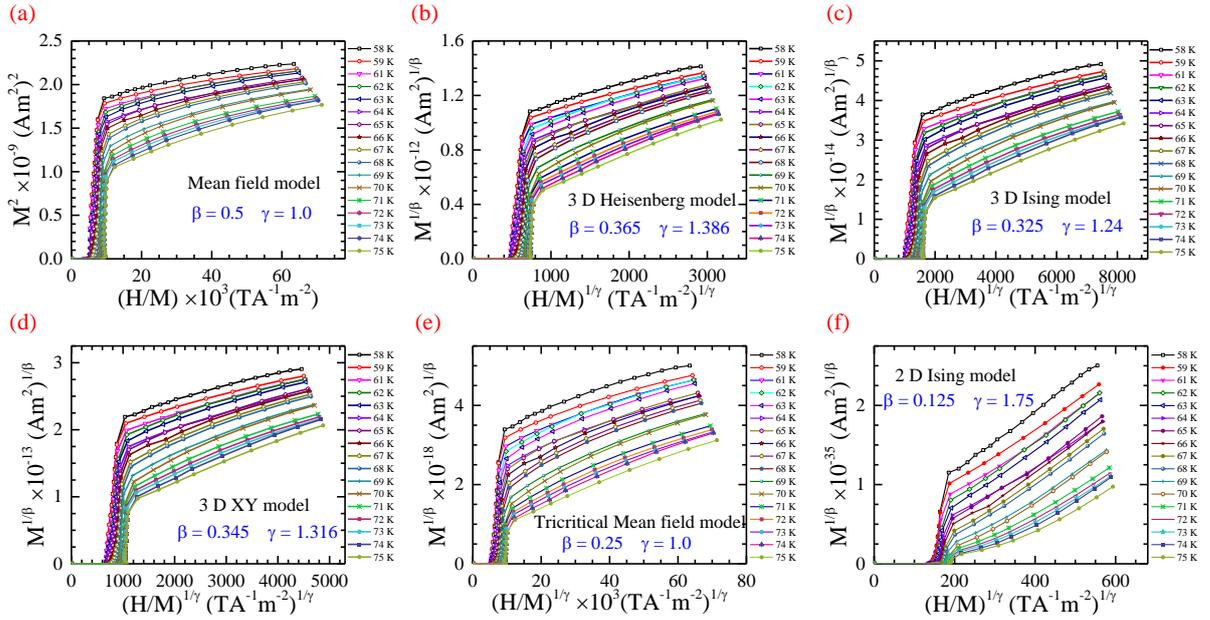

FIG. S6 MAP for isotherms in the critical regime for bulk CGT with (a) Mean field model (b) 3D Heisenberg model (c) 3D Ising model (d) 3D XY model (e) Tricritical mean field model and (f) 2D Ising model

## SI-7: Determination of critical exponents from dc and ac magnetization measurements

Using $T_c$ as a fitting parameter, the critical regime is located between (63.68 ±1.99) K to (66.59 ± 0.27) K (see Fig. 2(d) in MS). The critical exponent $\delta$ is determined to be 11.37(9) using $M = AH^{1/\delta}$ fit of the $M(H)$ curve at $T_c = 67$ K (Fig. S7 (a)). Inset of Fig. S7(a) shows $M(H)$ at 67 K in log-log scale. Using $\beta = 0.14$, $\gamma = 1.45$, $\delta$ is again independently estimated to be 11.35(7) using the Widom's scaling relation : $\delta = 1 + \frac{\gamma}{\beta}$. These two estimations of $\delta$ are in close agreement.

We did an independent estimation of $\gamma$ using zero field ac susceptibility data ($\chi_{ac}$), measured in a small oscillatory field of 3.8 Oe (ZFC warming) with a frequency of 211 Hz applied along sample's easy axis, $H \perp ab$. It has been extensively discussed in relevant literature that low-field (or better, zero field) data gives a more appropriate estimate of critical properties[18]. Our measurement shows a distinct hump in $Re\,(\chi_{ac})$ close to PM-FM transition, and the data was fitted in this neighborhood with $\chi = \chi_0 t^{-\gamma}$ (Fig. S7(b)) which gives the critical exponent $\gamma$ as (1.34 ± 0.08) and $T_c$ of (63.72 ± 0.38) K. These estimation of $T_c$, $\gamma$ and $\delta$ are in reasonable agreement with the findings of MAP and critical scaling analysis, and therefore, are intrinsic to the PM-FM phase transition in bulk CGT.

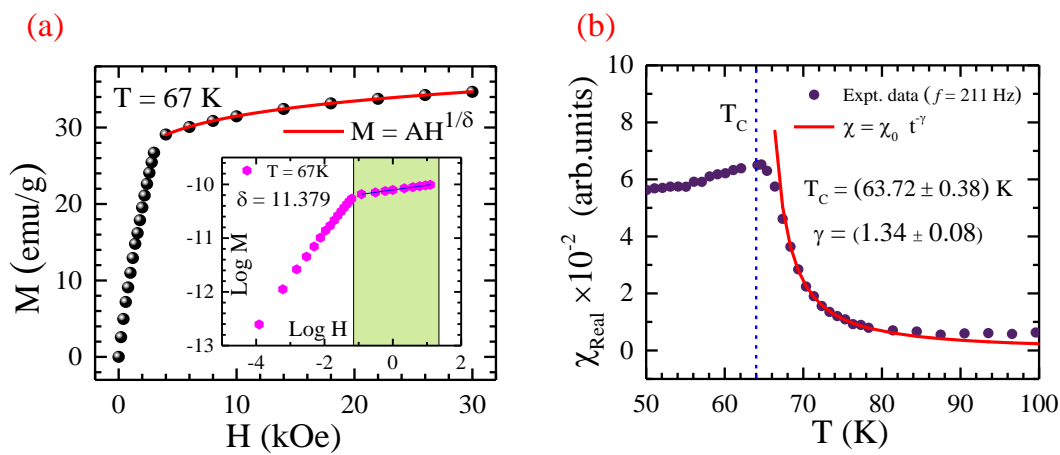

FIG. S7(a) $M(H)$ isotherm at $T_c = 67$ K fitted with $M = AH^{1/\delta}$. (inset) Log M vs Log H at $T_c$ with a straight line fit to determine $\delta$ (b) Real part of ZFC warming ac magnetic susceptibility of bulk CGT at 211 Hz frequency. The red solid curve is a fit with $\chi = \chi_0 t^{-\gamma}$ to extract $\gamma$ and $T_c$.

## SI-8: Details of theoretical calculations

### SI-8(A): Details of DFT calculations

First principles DFT calculations were performed using the Vienna *ab initio* simulation package (VASP)[19,20]. Electronic wave functions were expressed in terms of a plane wave basis set with an energy cutoff of 500 eV. Projector augmented wave (PAW)[21] potentials were used to represent the interaction between the valence electrons and the ion cores. The generalised gradient approximation (GGA) as proposed by Perdew, Burke and Ernzerhof (PBE)[22] was used to treat the exchange-correlation energy. The GGA+U[23] method was used to treat the strong correlations of Cr 3d electrons. It was previously reported[9] that the appropriate range of U should be within 0.2 < U < 1.7 eV in order to reproduce the correct experimental magnetic ground state. Hence U = 1eV was taken in our calculation. DFT-D3 method of Grimme[24] was used to treat the van der Waals interactions between the CGT layers. The structures were optimized by changing both the ionic positions and lattice parameters of the simulation cell until the energy and force on each atom converged to less than $10^{-7}$ eV and $10^{-6}$ eV/Å, respectively. For bilayer and monolayer a vacuum greater than 20 Å was used in order to avoid interactions between periodic images. Magnetocrystalline anisotropy energy (MAE) is calculated with the inclusion of spin-orbit coupling by taking the energy difference $E_a - E_c$ where $E_a(E_c)$ is the energy of the system when the spins orient along crystallographic $a(c)$ direction. Γ-centered Monkhorst-Pack k-point mesh[25] was used to perform Brillouin zone integrations. The density of k-point mesh was set at 0.15 Å$^{-1}$ during self-consistency calculations, and for MAE calculations a denser k-mesh of density 0.10 Å$^{-1}$ was used.

SI-8(B): Details of Dipolar Anisotropy Energy (DAE) calculations

Since the moments predominantly come from the Cr atoms, we consider the lattice spanned by these atoms in order to calculate DAE. Dipolar interaction energy of a central moment with all other moments on the lattice up to a cut off radius was calculated using the usual expression for interaction between two magnetic dipole moments separated by a distance $r$ as given in eqn. (S5). The cut off radius was chosen so that the interaction energy converged within 0.01 $\mu$eV for bulk and the bilayer and 0.001 $\mu$eV for the monolayer.

$$U = \frac{\mu_0}{4\pi}\frac{1}{r^3}[\vec{m_1}\cdot\vec{m_2} - 3(\vec{m_1}\cdot\hat{r})(\vec{m_2}\cdot\hat{r})] \text{ ------------------ (S5)}$$

where $\vec{m_1}$ and $\vec{m_2}$ are the interacting dipole moments.

Dipolar interaction energy was calculated for both in-plane and out-of-plane orientations of the Cr moments. Magnitude of all the moments were taken as $3\mu_B$ as obtained from DFT calculations. DAE is extracted as the difference in energies in in-plane and out-of-plane orientations of the moments. Details of these calculations is shown in Fig. S8.

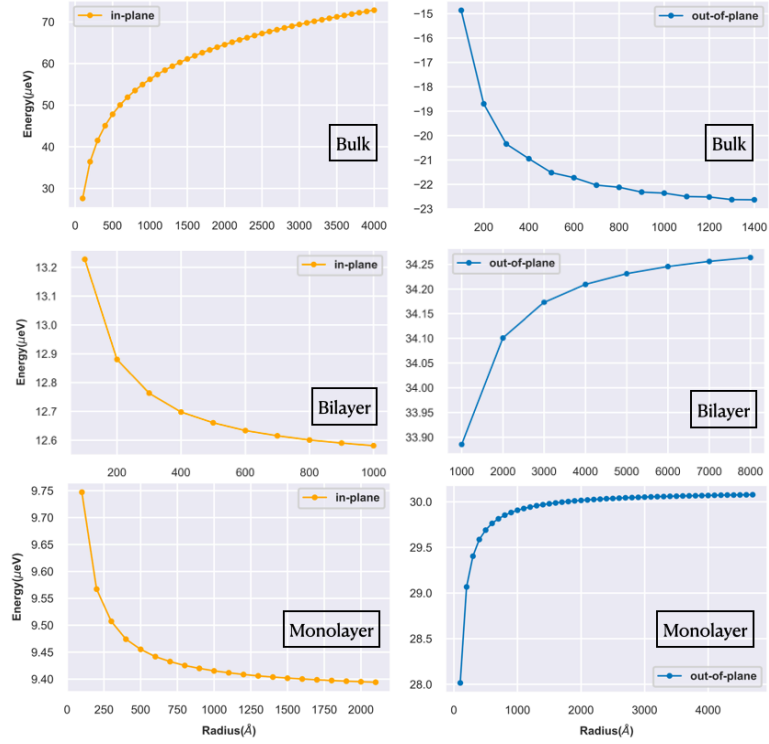

FIG. S8: Performing the real space dipolar sum over a sphere of certain radius for bulk, bilayer and monolayer CGT in in-plane and out-of-plane spin orientations.

## SI-9: Anisotropic consideration in magnon dispersion

In a simple spin wave picture, the dispersion relation of magnons in a ferromagnetic material at sufficiently low temperature (or, in other words, long-wavelength limit) is given by $E(q) \sim (1 - \cos(qa))$ [26], often approximated as $E(q) \sim q^2$ for small $q$. The dispersion is gapless and the corresponding magnetization is given by

$M(T) = M(0)(1 - bT^{3/2})$ ------- (S6, Bloch's law).

This treatment, however, does not consider the effect of anisotropy. Note that magnetic anisotropy is critically important for long-range ordering (or, spontaneous symmetry breaking) in two dimensional materials as, for a 2D *isotropic* system, long-wavelength Goldstone excitations can be generated with vanishingly small energy cost and hence disrupts the ordering. The effect of incorporating an anisotropy term is to introduce an excitation gap ($\Delta$) in the dispersion spectrum (which, for isotropic magnetic systems, is close to zero[27]). The modified dispersion relation is given by $E(q) = \Delta + Dq^2$ and the resulting magnetization as a function of $T$ is given by[28]

$M(T, H) = M(0, H) - g\mu_B \left(\frac{k_B T}{4\pi D}\right)^{\frac{3}{2}} f_{\frac{3}{2}}\left(\frac{\Delta'}{k_B T}\right)$ -------------- (S7)

where $D$ is the spin stiffness co-efficient, $\Delta' = \Delta + g\mu_B H$ and $f_{\frac{3}{2}}(y) = \sum_{n=1}^{\infty} \frac{\exp(-ny)}{n^{3/2}}$ is the Bose-Einstein integral function. In order to find out the effect of anisotropy in CGT, in Fig. S9(a), we fit the $M_S(T)$ data for bulk single crystal of CGT at low $T$ ($T < 0.4\, T_c$) with the gapless dispersion (eqn.(S6)) and gapped dispersion (eqn.(S7)).

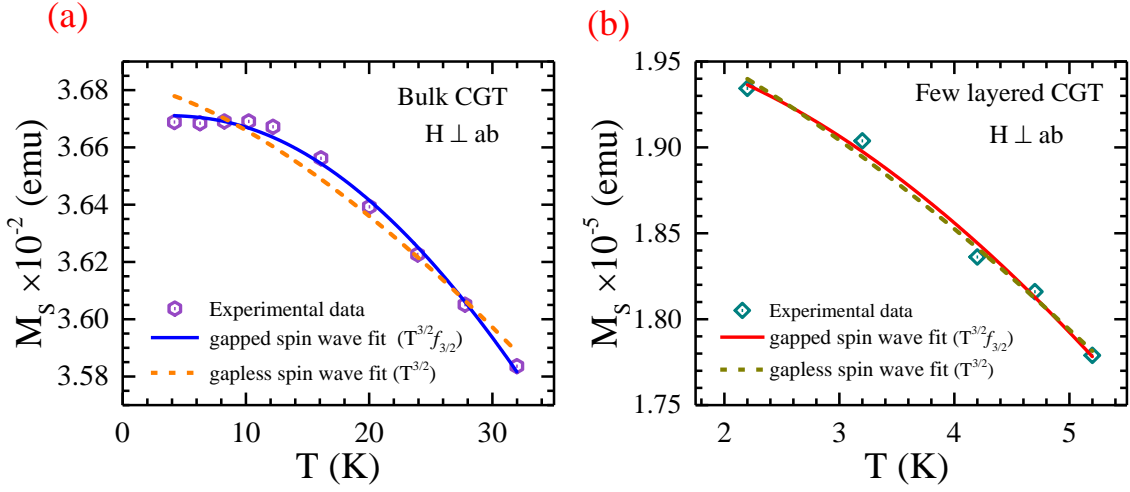

FIG. S9 (a) Temperature dependence of spontaneous magnetization for CGT bulk crystal at low $T$ ($T < 0.4\, T_c$) obtained from $H \perp ab$ data with fits to gapless spin wave dispersion (eqn. S6) in orange dashed curve and fit to gapped spin wave dispersion (eqn. S7) in blue solid curve. (b) Temperature dependence of spontaneous magnetization for few-layered CGT at low $T$ for $H \perp ab$ with fits to gapless spin wave dispersion (eqn. S6) in yellow dashed curve and fit to gapped spin wave dispersion (eqn. S7) in red solid curve.

It is clear that the gapless equation (S6) is not a good fit to the experimental data, necessitating the importance of taking anisotropy (i.e. eqn. S7) into consideration. Following are the extracted parameters from fitting with the gapped magnon dispersion spectrum (eqn. S(7))

$D = (26.09 \pm 0.02)$ meVÅ$^2$ and $\Delta = (1.38 \pm 0.27)$ meV

The spin stiffness co-efficient ($D$) is in reasonable agreement with reported values[29] for bulk CGT. Anisotropy-induced gap $\Delta$ is slightly higher than reported values, presumably because of using comparatively broader $T$-window for fitting and/or high sensitivity of Bose-Einstein integral function to the excitation gap value $\Delta$. Next, we investigate the effect of anisotropy in low dimensional limit. Fitting of $M_S(T)$ data for $T < 0.35 T_c'$ (i.e. $T < 5.2$ K in Fig. 3(d) in MS) with eqn. (S7) gives a value of $\Delta = (0.14 \pm 0.03)$ meV (see Fig. S9(b)), which is an order of magnitude smaller than its bulk counterpart. The observed drop in $\Delta$ strongly suggests significant weakening of anisotropy with lowered dimensionality.